\documentclass[epjCONF]{svjour}
\usepackage{graphics}
\usepackage[varg]{txfonts} 
\usepackage[latin1]{inputenc}
\newcommand{\lamost}{{\sc lamost}}
\newcommand{\standard}{{\sc standard}}
\newcommand{\kasc}{{\sc kasc}}
\newcommand{\planet}{{\sc planet}}
\newcommand{\extra}{{\sc extra}}
\newcommand{\field}{{\sc field}}
\newcommand{\kepler}{{\it Kepler}}
\newcommand{\lk}{{\sc lamost}-{\it Kepler}}
\newcommand{\degree}{$^{\circ}$}
\newcommand{\teff}{$T_{\rm eff}$}
\newcommand{\logg}{$\log g$}
\newcommand{\vsini}{$v \sin i$}

\newcommand{\feh}{$[Fe/H]$}
\newcommand{\pointing}{LK-field}
\newcommand{\pointings}{LK-fields}
\newcommand{\project}{LK-project}
\newcommand{\Kp}{$K\rm p$}
\newcommand{\kic}{KIC\,10}
\newcommand{\vrad}{$v_{\rm rad}$}

\newcommand{\ulyss}{{\sc ul}y{\sc ss}}
\newcommand{\rotfit}{{\sc rotfit}}
\newcommand{\mkclass}{{\sc mkclass}}
\newcommand{\kms}{km\,s$^{-1}$}
\session-title{CoRoT Symposium 3, Kepler KASC-7 joint meeting: The Space Photometry Revolution}
\begin{document}
\title{{\sc lamost} observations in the {\it Kepler} field}
\author{P. De Cat\inst{1}\fnmsep\thanks{\email{Peter.DeCat@oma.be}} \and 
J.N. Fu \inst{2} \and
X.H. Yang \inst{2} \and
A.B. Ren \inst{2} \and
A. Frasca \inst{3} \and 
J. Molenda-\.Zakowicz \inst{4} \and
G. Catanzaro \inst{3} \and 
R.O. Gray \inst{5} \and
C.J. Corbally \inst{6} \and 
J.R. Shi \inst{7} \and 
H.T. Zhang \inst{7} \and 
A.L. Luo \inst{7}
}
\institute{Royal Observatory of Belgium, Brussels, Belgium \and
Beijing Normal University, Beijing, China \and
INAF (Osservatorio Astrofisico di Catania), Catania, Italy \and
Instytut Astronomiczny, Uniwersytet Wroc{\l}awski,  Wroc{\l}aw, Poland \and
Appalachian State University, Boone, North Carolina (USA) \and
Vatican Observatory Research Group, University of Arizona, Tucson, Arizona (USA) \and
Key Lab for Optical Astronomy, National Astronomical Observatories, Chinese Academy of Sciences, Beijing 100012, China 
}
\abstract{
The Large Sky Area Multi-Object Fiber Spectroscopic Telescope (\lamost) at the Xinglong observatory in China is a new 4-m telescope equipped with 4,000 optical fibers. 
In 2010, we initiated the \lk\ project. 
We requested to observe the full field-of-view of the nominal \kepler\ mission with the \lamost\ to collect low-resolution spectra for as many objects from the \kic\ catalogue as possible. 
So far, 12 of the 14 requested \lamost\ fields have been observed resulting in more than 68,000 low-resolution spectra. 
Our preliminary results show that the stellar parameters derived from the \lamost\ spectra are in good agreement with those found in the literature based on high-resolution spectroscopy.
The \lamost\ data allows to distinguish dwarfs from giants and can provide the projected rotational velocity for very fast rotators.
}
\maketitle
\section{Introduction}
\label{intro}
The space mission \kepler\ has been designed to detect Earth-like planets around solar-type stars by the transit method \cite{Koch2010ApJ...713L..79K}. 
It was launched on 2009 March 7 at 03:49:57.465 UTC and has been collecting ultraprecise photometry with a spectral bandpass from 400~nm to 850~nm for a fixed field-of-view (FoV) of 105 square degrees in the constellations Lyra and Cygnus. 
In May 2013, a second of the four reaction wheels of the  \kepler\ spacecraft failed, preving the telescope from precisely pointing toward stars. 
Even though ultrahigh precision photometry can no longer be collected, the project is continued as the K2 mission and the legacy of the \kepler\ photometry is a pure goldmine for asteroseismic studies of all types of pulsating stars.

The success of asteroseismic studies has been shown to depend crucially on the availability of basic stellar parameters, such as the effective temperature (\teff), surface gravity (\logg), metallicity (\feh), and the stellar rotation rate (\vsini) (e.g. \cite{Cunha2007A&ARv..14..217C}).
These parameters can not be derived from the \kepler\ data as is the case for data with multi-colour photometry or spectroscopy. 
Before the launch of the \kepler\ spacecraft, there was a large effort to derive the stellar parameters from Sloan photometry for potential \kepler\ targets. 
These are available in the \kepler\ Input Catalogue (\kic; \cite{Latham2005AAS...20711013L}). 
Unfortunately, not for all stars \kic\ stellar parameters are available.
Moreover, the precision of the \teff\ and \logg\ in \kic\ is generally too low for asteroseismic modelling, especially for hot and peculiar stars (e.g. \cite{McNamara2012AJ....143..101M}). 
Also, additional information on the stellar chemical composition and rotation rate is lacking. Hence, to exploit the \kepler\ data best, additional ground-based {\it spectroscopic} data are required (e.g. \cite{Uytterhoeven2010AN....331..993U}). 

The Large Sky Area Multi-Object Fiber Spectroscopic Telescope (also called the Guo Shou Jing Telescope) is a unique astronomical instrument located at the Xinglong observatory (China) that combines a large aperture (4-m telescope) with a wide FoV (circular with diameter 5\degree) \cite{Wang1996ApOpt..35.5155W}.
The focal surface is covered with 4,000 optical fibers (Fig.\,\ref{fig:LKfields}, left) connected to 16 sets of multi-objective optical spectrometers with 250 optical fibers each \cite{Xing1998SPIE.3352..839X}.
Each spectrometer has two CCD cameras to obtain low resolution spectra ($R \simeq 1800$) in two wavelength regions (blue arm: $370-590$\,nm; red arm: $570-900$\,nm). 
The active optics technique is used to control the reflecting corrector \cite{Su1998SPIE.3352...76S}. 
The \lamost\ has a quasi-meridian transit configuration capable of tracking the motion of celestial objects during about 4\,hours while they are passing the meridian. 
Hence, the \lamost\ is an ideal instrument to perform spectroscopic follow-up for targets of the \kepler\ mission in an efficient way as it is capable of collecting low-resolution spectra for thousands of objects down to magnitude 17.8 simultaneously.
For more detailed information about the \lamost, see \cite{Cui2012RAA....12.1197C} and \cite{Zhao2012RAA....12..723Z}.

\section{The \lamost-\kepler\ project}
\label{sec:LKproject}

\begin{figure}
\resizebox{0.41\columnwidth}{!}{\includegraphics{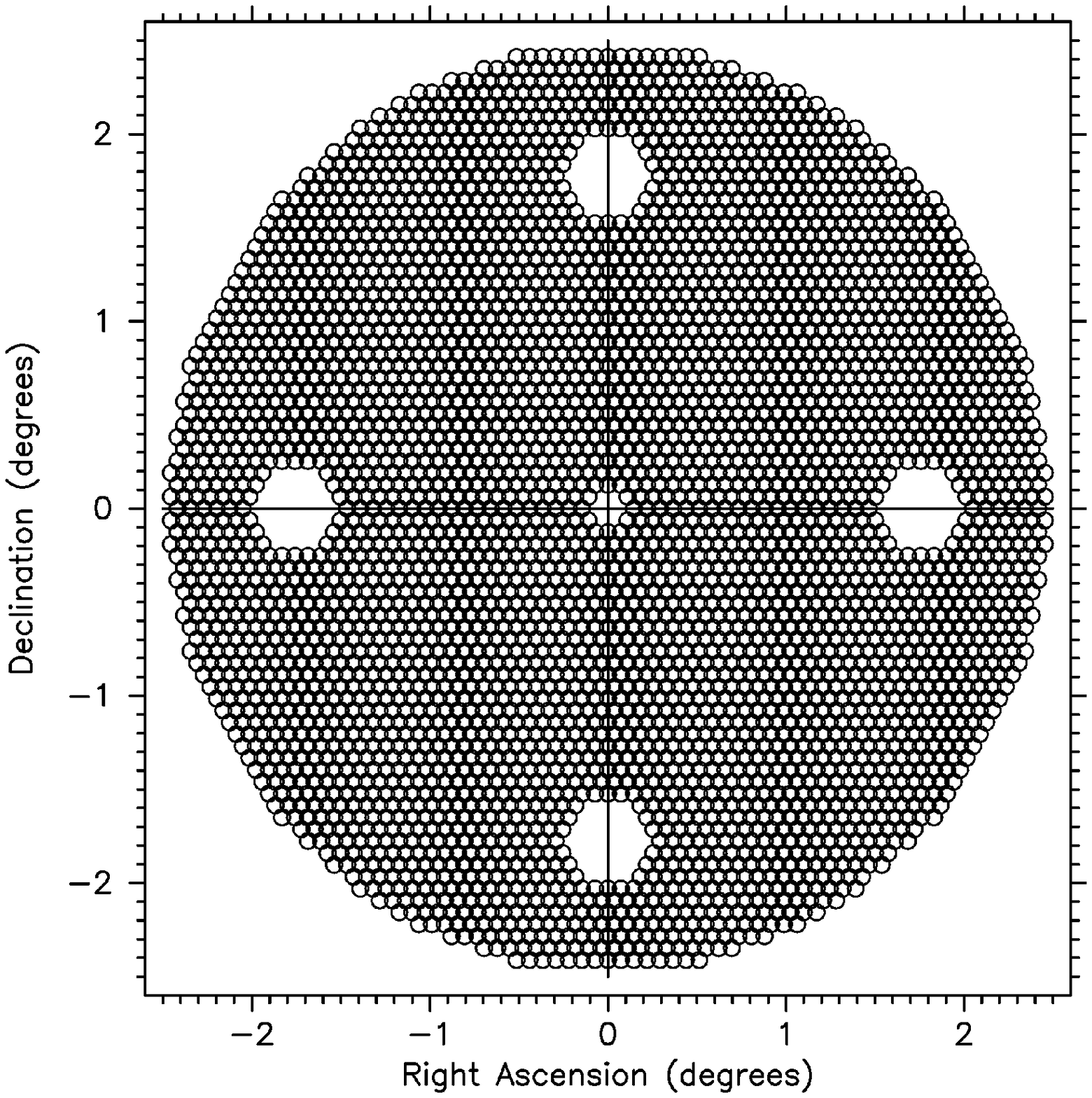} }
\resizebox{0.57\columnwidth}{!}{\includegraphics{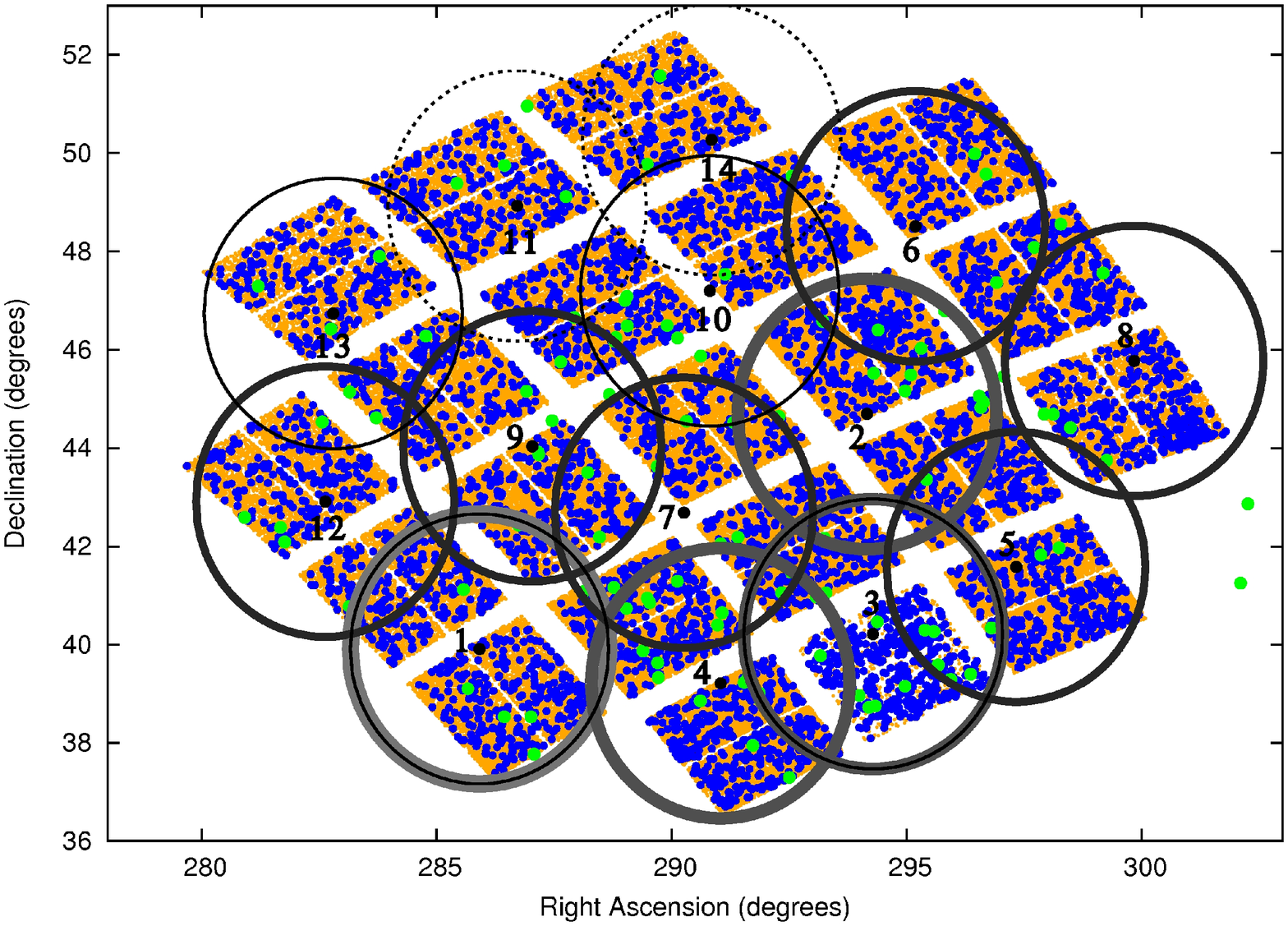} }
\caption{
{\bf Left:} 
The 4,000 fibers are homogeneously distributed on the focal plane except for the position of active optics wavefront sensor (central hole) and 4 Guiding CCD cameras (off-centre holes).
{\bf Right:} 
Representation of the targets of scientific interest in the FoV of the \kepler\ mission. 
The black dots refer to the centres of the 14 \pointings\ that cover the \kepler\ FoV (cf. Table\,\ref{tab:LKfields}). 
We use green for \standard\ targets, blue for \kasc\ targets and orange for \planet\ targets. 
The scientific importance is reflected in the size of the symbols. 
The \pointings\ that have been observed in 2011, 2012, 2013, and 2014 are indicated by the circles drawn with a full line going from thick to thin and from grey to black, respectively. 
The two fields that remain unobserved so far are given with a black dashed line.
(colour representation only available online)
}
\label{fig:LKfields}
\end{figure}

\begin{table}
\caption{
For each \pointing, we give the right ascension \& declination of the central bright star and the name of the open cluster that it contains (if applicable). 
For those that have already been observed up to June 2014, we additionally give the date of observation, the number of plates that were used to observe the \pointing, the total number of \lamost\ spectra and the number of different \kepler\ targets (\standard\ + \kasc\ + \planet\ + \extra) and field targets that were observed. 
}
\label{tab:LKfields}
\begin{tabular}{lllcccrrrrrrr}
\hline\noalign{\smallskip}
    \pointing& RA(2000)      & DE(2000)      & Cluster & Date       &\# &  Spectra & \kepler\ &   Field \\
\hline\noalign{\smallskip}
    LK01     & 19:03:39.258  & +39:54:39.24  &         & 30/05/2011 & 2 &    1,107 &   1,107  &       0 \\        
             &               &               &         & 08/06/2011 & 2 &      966 &     963  &       3 \\
             &               &               &         & 02/06/2014 & 1 &    3,248 &   3,246  &       2 \\
    LK02     & 19:36:37.977  & +44:41:41.77  & NGC6811 & 04/06/2012 & 1 &      563 &     462  &     101 \\        
    LK03     & 19:24:09.919  & +39:12:42.00  & NGC6791 & 15/06/2012 & 3 &    7,602 &   7,531  &      71 \\    
    LK04     & 19:37:08.862  & +40:12:49.63  & NGC6819 & 17/06/2012 & 3 &    9,629 &   9,492  &     137 \\ 
    LK05     & 19:49:18.139  & +41:34:56.85  &         & 05/10/2013 & 2 &    6,400 &   6,323  &      77 \\ 
             &               &               &         & 22/05/2014 & 1 &    1,525 &   1,525  &       0 \\
    LK06     & 19:40:45.383  & +48:30:45.10  &         & 22/05/2013 & 1 &      394 &     394  &       0 \\
             &               &               &         & 23/05/2013 & 1 &    1,834 &   1,834  &       0 \\
             &               &               &         & 14/09/2013 & 1 &    2,823 &   2,807  &      16 \\
    LK07     & 19:21:02.816  & +42:41:13.07  &         & 19/05/2013 & 1 &    1,936 &   1,936  &       0 \\ 
             &               &               &         & 26/09/2013 & 1 &    3,004 &   2,955  &      49 \\ 
    LK08     & 19:59:20.425  & +45:46:21.15  & NGC6866 & 02/10/2013 & 1 &    2,807 &   2,771  &      36 \\
             &               &               &         & 17/10/2013 & 1 &    2,592 &   2,565  &      27 \\
             &               &               &         & 25/09/2013 & 2 &    5,574 &   5,521  &      53 \\
             &               &               &         & 25/10/2013 & 1 &    2,791 &   2,764  &      27 \\
    LK09     & 19:08:08.340  & +44:02:10.88  &         & 04/10/2013 & 1 &    3,118 &   3,116  &       2 \\
    LK10     & 19:23:14.829  & +47:11:44.80  &         & 20/05/2014 & 2 &    2,389 &   2,388  &       1 \\
    LK11     & 19:06:51.499  & +48:55:31.77  &         &            &   &          &          &         \\
    LK12     & 18:50:31.041  & +42:54:43.72  &         & 07/10/2013 & 1 &    2,741 &   2,711  &      30 \\ 
    LK13     & 18:51:11.993  & +46:44:17.52  &         & 02/05/2014 & 1 &    1,968 &   1,965  &       3 \\
             &               &               &         & 29/05/2014 & 2 &    3,715 &   3,707  &       8 \\
    LK14     & 19:23:23.787  & +50:16:16.64  &         &            &   &          &          &         \\ 
\noalign{\smallskip}\hline
\end{tabular}
\end{table}

In 2010, we initiated the \lamost-\kepler\ project (\project) to observe as many objects in the \kepler\ FoV as possible from the start of the test phase of the \lamost\ onwards. 
The main goal is to determine both the stellar parameters and the spectral class of the observed objects in a homogeneous way. 
Moreover, with low-resolution spectra it is possible to estimate the radial velocity (\vrad) and the \vsini\ (to detect rapid rotation) of the observed objects.

We have composed a prioritised target list consisting of, from high to low priority, $\sim$250 \standard\ targets (MK secondary standard stars; originally selected for calibration purposes), $\sim$7,000 \kasc\ targets (scientific interest for the \kepler\ Asteroseismic Science Consortium), $\sim$150,000 \planet\ targets (scientific interest for the planet search group), $\sim$1,000,000 \extra\ targets (other targets from \kic\ catalogue; no specific scientific interest), and \field\ targets (objects from the USNO-B catalogue \cite{Monet2003AJ....125..984M}; to fill the fiber holes).
As \lamost\ observations can only be optimized for objects with a magnitude interval of $\sim$5 magnitudes, the targets were subdivided into a bright (9\,$<$\,\Kp\,$\leq$14) and faint (14\,$<$\,\Kp) group.
Highest priority was given to objects without \kic\ parameters.
As the temperature determination is the least accurate for hot stars, the others were sorted from high to low \kic\ \teff.
The bright targets were additionally sorted from faint to bright (to avoid saturation + brightest objects can be observed with smaller telescopes) and the faint ones from bright to faint (to avoid underexposure).

We selected 14 \lk\ fields (\pointings) to cover the \kepler\ FoV (Fig.\,\ref{fig:LKfields}, right).
Each \pointing\ contains a central bright star ($V < 8$) for the active optics and four fainter stars ($V < 17$) in the off-centre holes for the guiding of the CCD cameras (cf. Fig.\,\ref{fig:LKfields}, left).
A new system for the positioning of the 4,000 fibers of the \lamost\ has been developed \cite{Xing1998SPIE.3352..839X} and the code ``Survey Strategy System'' is used to prepare the observation plans of the \project\ in the most efficient way.
More details about the requested \pointings\ are given in Table\,\ref{tab:LKfields}.
Up to June 2014, all but two of the \pointings\ have been observed.

\section{Preliminary results}
\label{sec:LKresults}

There are three teams working on the analysis of the full set of \lamost\ spectra, each using their own methods.

The Asian team is determining \teff, \logg, \feh\ and \vrad\ with a version of the software package \ulyss\ \cite{Koleva2009A&A...501.1269K} adapted to the \lamost\ data (Ren et al., in preparation).
It performs a $\chi^2$ minimization for a fit of a parametric model to the observed spectra in pixel space. 
The European team is using an adapted version of the code \rotfit\ (e.g. \cite{Frasca2003A&A...405..149F}) for the same purposes (Frasca et al., in preparation). 
The observed spectra are fitted to those available in a grid of high-resolution spectra for a selection of more than 1,000 comparison stars with known stellar parameters from the Indo-U.S. Library of Coud\'e Feed Stellar Spectra \cite{Valdes2004ApJS..152..251V} after degrading them to the resolution of the \lamost\ spectra. 
As the library spectra are in the laboratory rest frame and are corrected for their heliocentric radial velocity, they also serve as templates to derive \vrad\ with the cross-correlation technique. 
Moreover, this analysis method is also capable of giving a rough estimation of \vsini\ for rapidly rotating stars. 
The resulting stellar parameters for \kasc\ objects are compared to those derived from available ground-based follow-up spectroscopy obtained with other instruments to check the consistency of the results.
\teff\ and \vrad\ can be determined with an accuracy of $\sim$4\% and $\sim$13\,\kms, respectively, while \logg\ can only distinguish between main-sequence and evolved stars.
\feh\ is in general consistent with solar metallicities.
Estimates for \vsini\ are only relibable for stars rotating faster than 150\,\kms.
 
The American team has developed the code \mkclass\ for classifying stars automatically on the MK spectral classification system independent of the stellar parameter determination \cite{Gray2014AJ....147...80G}. 
This method requires a library of spectral standards and is designed to classify stellar spectra by direct comparison with MK standards using the same criteria as human classifiers. 
Moreover, \mkclass\ is capable of recognizing many of the common spectral peculiarities.
For the \lamost\ classifications, the flux-calibrated standards library with 3.6\AA-resolution spectra obtained using the GM spectrograph at the Dark Sky Observatory of Appalachian State University is employed. 
Therefore, the \lamost\ spectral resolution is slightly degraded to match that of the spectral library. 
The accuracy of the resulting classifications does not depend upon the accuracy of the flux calibration of the \lamost\ spectra. 
Based on tests on spectra classified by humans, the systematic error and standard deviation of the spectral and luminosity classes are 0.1 and 0.5 spectral subclasses (where a unit spectral subclass is the difference between, for instance, F5 and F6) and 0.02 and 0.5 luminosity classes (where a unit luminosity class is the difference between, for instance, a dwarf (V) and subgiant (IV) classification), respectively. 
Thus the accuracy of \mkclass\ is similar to the level of agreement obtained by two independently working, skilled human classifiers.

\section{Conclusions}
\label{sec:LKconclusions}

The \project\ is an ambitious observational project that is providing accurate stellar parameters derived from low resolution spectra in an efficient way (4,000 fibers) for objects fainter than most ground-based facilities allow (4-m telescope).
The project is not finished yet: two of the requested \pointings\ were not observed (scheduled before the end 2014) and not all of \lamost\ spectra have been analysed by the different teams (work in progress).
For a full description of the project and its first results, we refer to De Cat et al. (to be submitted to ApJS).
The \lamost\ spectra are available upon request (Peter.DeCat@oma.be).

\end{document}